\documentclass[12pt]{article}
\usepackage{bbm}
\usepackage{latexsym}
\usepackage{mathrsfs}
\usepackage{graphicx}
\usepackage{subfigure}
\usepackage{amssymb}
\usepackage{amsmath}
\usepackage{amsthm}
\usepackage{color}
\usepackage{cite}
\usepackage{float}
\usepackage{indentfirst}
\usepackage{anysize}\marginsize{45mm}{45mm}{40mm}{50mm}
\usepackage{caption}
\usepackage{float}
\usepackage{multirow}

\usepackage{enumerate}

\usepackage{latexsym, bm}
\newtheoremstyle{lemma}{\topsep}{\topsep}%
     {}
     {}
     {\bfseries}
     {}
     {0.1em}
     {\thmname{#1}\thmnumber{ #2}\thmnote{ #3}}
\theoremstyle{lemma}  

\newtheorem{theorem}{Theorem}[section]    

\newtheorem{corollary}[theorem]{Corollary}

\numberwithin{equation}{section}

\title{On complexity of substructure connectivity and restricted connectivity of graphs\thanks{This research was partially supported by the National Natural Science Foundation of China (Nos. 11801061 and 12261071), NSF of Qinghai Province (No. 2025-ZJ-902T) }}

\author{Huazhong L\"{u}$^{1}$\thanks{Corresponding author.} and Tingzeng Wu$^{2}$\\
{\small $^{1}$School of Mathematical Sciences, } \\
{\small University of Electronic Science and Technology of China,} \\
{\small Chengdu, Sichuan 610054, P.R. China}\\
{\small E-mail: lvhz@uestc.edu.cn}\\
{\small $^{2}$School of Mathematics and Statistics, Qinghai Nationalities University, }\\
{\small Xining, Qinghai 810007, P.R. China} \\
{\small E-mail: mathtzwu@163.com}\\}

\date{}
\begin{document}

\maketitle
\begin{abstract}
The connectivity of a graph is an important parameter to evaluate its reliability. $k$-restricted connectivity (resp. $R^h$-restricted connectivity) of a graph $G$ is the minimum cardinality of a set $S$ of vertices in $G$, if exists, whose deletion disconnects $G$ and leaves each component of $G-S$ with more than $k$ vertices (resp. $\delta(G-S)\geq h$). In contrast, structure (substructure) connectivity of $G$ is defined as the minimum number of vertex-disjoint subgraphs whose deletion disconnects $G$. As generalizations of the concept of connectivity, structure (substructure) connectivity, restricted connectivity and $R^h$-restricted connectivity have been extensively studied from the combinatorial point of view. Very little is known about the computational complexity of these variants, except for the recently established NP-completeness of $k$-restricted edge-connectivity. In this paper, we prove that the problems of determining structure, substructure, restricted, and $R^h$-restricted connectivity are all NP-complete.

\vskip 0.1 in


\end{abstract}

\section{Introduction}

Throughout this paper, we consider simple and undirected graphs. Let $G=(V,E)$ be a graph, where $V$ is the vertex-set of $G$ and $E$ is the edge-set of $G$. The degree of a vertex $v$ is the number of incident edges, denoted by $d_G(v)$ or simply $d(v)$ when the context is clear. The minimum degree of $G$ is $\delta(G)$ and the maximum degree is $\Delta(G)$. For any subset $X\subset V$, the {\em closed neighborhood} of $X$ is defined to be all neighbors of any vertex $x\in X$ together with $X$, denoted by $N[X]$, while the {\em open neighborhood} of $X$ is $N[X]\setminus X$, denoted by $N(X)$. If $X=\{x\}$, then we write $N[x]$ and $N(x)$, respectively. The subgraph induced by $X$ is denoted by $G[X]$. For other standard graph-theoretical terminologies and notations not defined here, we follow Bondy \cite{Bondy}.

The connectivity and edge-connectivity of a graph play important roles in measuring network reliability and have been widely studied in the literature from a combinatorial point of view. Generally, the higher the connectivity, the more reliable the network is. With the rapid development of multiprocessor systems, it is meaningful and necessary to accurately evaluate their reliability. However, the classical connectivity and edge-connectivity always underestimate the resilience of large networks as these two parameters tacitly assume that all vertices adjacent to, or all edges incident to, the same vertex can potentially fail simultaneously. This is highly unlikely in large networks.

To address the deficiencies of connectivity and edge-connectivity, several generalizations were introduced by graph theorists and computer scientists. In the seminal paper \cite{Harary}, Harary introduced the concept of conditional connectivity to evaluate the reliability of a graph by requiring that, if the graph is disconnected after faults, each resulting component must possess a given property $P$. As a kind of conditional connectivity, in 1988, Esfahanian and Hakimi \cite{Esfahanian} proposed the definition of {\em restricted edge-connectivity}. An edge-cut $S\subseteq E(G)$ is called a {\em restricted edge-cut} if there exists no isolated vertices in $G-S$. The {\em restricted edge-connectivity} is the minimum cardinality over all restricted edge-cuts $S$. Motivated by this, F\`{a}brega and Fiol \cite{Fabrega1,Fabrega2} generalized it to the concept of $k$-restricted edge-connectivity (where $k$ is a nonnegative integer) as follows. Given a graph $G$ and a non-negative integer $k$, the {\em $k$-restricted connectivity} (resp. {\em $k$-restricted edge-connectivity}) of $G$, denoted by $\kappa_k(G)$ (resp. $\lambda_k(G)$), is the minimum cardinality of a set of vertices (resp. edges) in $G$, if exists, whose deletion disconnects $G$ and leaves each remaining component of $G$ with more than $k$ vertices. In particular, a connected graph $G$ is called {\em $\lambda_k$-connected} if $\lambda_k(G)$ exists. Moreover, if $\lambda_k(G)$ exists, then $\lambda_i(G)$ exists for any positive integer $i<k$ and $\lambda_i(G)\leq\lambda_k(G)$. Thereafter, Latifi et al. \cite{Latifi} further generalized this measurement to $R^h$-restricted (edge-)connectivity by requiring each vertex to have at least $h$ fault-free neighbors. The $R^h$-restricted (edge-)connectivity of $G$, denoted by $\kappa^h(G)$ (resp. $\lambda^h(G)$), is the minimum number of vertices (resp. edges) of $G$ whose deletion disconnects $G$ and each vertex in the remaining subgraph has degree at least $h$. Thus, the higher $k$-restricted or $R^h$-restricted (edge-)connectivity ensures that after the failure of a set of vertices (or edges), no component of the network is ``too small'', allowing each surviving component to maintain minimal functionality.


Recently, as another variant of traditional connectivity, Lin et al. \cite{Lin} introduced structure and substructure connectivity to evaluate the fault tolerance of a network from the perspective of a single vertex, as well as some special structures of the network. Let $\mathcal{F}=\{F_1,F_2,\cdots,F_t\}$ be a set of pairwise disjoint connected subgraphs of $G$ and let $V(\mathcal{F})=\mathop{\cup}\limits_{i=1}^{t}{V(F_i)}$. Then $\mathcal{F}$ is a {\em subgraph-cut} of $G$ provided that $G-V(\mathcal{F})$ is disconnected or trivial. Let $H$ be a connected subgraph of $G$, then $\mathcal{F}$ is an {\em $H$-structure-cut} if $\mathcal{F}$ is a subgraph-cut, and each element in $F$ is isomorphic to $H$. The {\em $H$-structure connectivity} of $G$, written $\kappa(G; H)$, is the minimum cardinality over all $H$-structure-cuts of $G$. Similarly, if $\mathcal{F}$ is a subgraph-cut and each element of $\mathcal{F}$ is isomorphic to a connected subgraph of $H$, then $\mathcal{F}$ is called an {\em $H$-substructure-cut}. The {\em $H$-substructure connectivity} of $G$, written $\kappa^s(G; H)$, is the minimum cardinality over all $H$-substructure-cuts of $G$.


The refined connectivity of graphs has been widely studied, for example, structure and substructure connectivity \cite{Ba,Lin,Lv2,Sabir,Lv,Lei,Zhao}, component connectivity \cite{Chang,Guo,Hsu,Li,Liu}, $k$-restricted connectivity \cite{Chang2,Li,Lu}, and $R^h$-restricted (edge-)connectivity \cite{Yang,Li2}. Cornaz et al. \cite{Cornaz} showed that the vertex $k$-cut problem, defined as finding a minimum-cardinality vertex set whose removal leaves at least $k$ components, is NP-hard for any fixed $k\geq3$. In 2017, Montejano and Sau \cite{Montejano} demonstrated that it is NP-hard to determine the exact value of $\lambda_k(G)$ even for $\lambda_k$-connected graphs, implying that determining $k$-restricted edge-connectivity of graphs is NP-hard. Additionally, Goldschmidt and Hochbaum \cite{Goldschmidt} proved that $k$-component edge-connectivity problem is NP-hard if $k$ is an input parameter but admits a polynomial time algorithm if $k$ is a constant. Thus, a natural question arises: what are the computational complexities of structure, substructure, $k$-restricted, and $R^h$-restricted connectivity in general graphs? In this paper, we study these problems.


The rest of the paper is organized as follows. In Section 2, we prove NP-completeness of the substructure connectivity by reducing dominating set problem to it. Then, as a corollary, we give NP-completeness of structure connectivity. In Sections 3, we show that $k$-restricted connectivity and $R^h$-restricted connectivity are NP-complete, respectively. Finally, we conclude this paper in Section 4.

\section{Substructure and structure connectivity}

We first present the decision problem of the substructure connectivity as follows.

\textsc{$H$-Substructure Connectivity}

\vskip 0.00 in

\textit{Instance:} Given a nonempty graph $G=(V,E)$, a subgraph $H$ of $G$ and a positive integer $k<|V|$.

\vskip 0.00 in

\textit{Question:} Is $\kappa^{s}(G,H)\leq k$?


A {\em dominating set} of $G$ is a subset $D\subseteq V$ such that every vertex not in $D$ is adjacent to one member of $D$. The {\em dominating number} is the number of vertices in a smallest dominating set. The decision version of the dominating set problem is: Given a graph $G=(V,E)$ and a positive integer $d\leq|V|$, is there a dominating set $D$ of size not greater than $d$ such that for each vertex $u\in V-D$ there is a vertex $v\in D$ with $uv\in E$? It is a classical NP-complete problem showed as problem GT2 in p. 190 \cite{Garey}. In fact, it is also stated that dominating set problem remains NP-complete for planar graphs with maximum degree three.

\begin{theorem}{\bf .}\label{substr}
The \textsc{$H$-Substructure Connectivity} is NP-complete when $H=K_{1,M}$ for any integer $M\geq3$.
\end{theorem}
\begin{proof}
Obviously, \textsc{$H$-Substructure Connectivity} is in NP, because we can check in polynomial time whether a set of disjoint connected subgraphs of $K_{1,M}$ is a substructure cut. It remains to show that \textsc{$H$-Substructure Connectivity} is NP-hard when $H=K_{1,M}$, $M\geq3$. We establish NP-hardness by a reduction from the dominating set problem for planar graphs with maximum degree three.

Given a connected planar graph $G=(V,E)$ with $\Delta(G)\leq3$, we construct a graph $G'=(V',E')$ from $G$ as follows (see Fig. \ref{g-NPC-substr}). We may assume that the order of $G$ is $n$.

\begin{figure}[h]
\centering
\includegraphics[height=85mm]{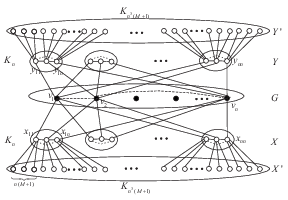}
\caption{The graph $G'$ constructed for proving NP-completeness of the substructure connectivity.}
\label{g-NPC-substr}
\end{figure}
Set
$$X=\{x_{ij}|1\leq i\leq n, 1\leq j\leq n\},$$
$$Y=\{y_{ij}|1\leq i\leq n, 1\leq j\leq n\},$$
$$X'=\{x'_{s}|1\leq s\leq n^3(M+1)\},$$
$$Y'=\{y'_{t}|1\leq t\leq n^3(M+1)\},$$
$$V=\{v_1,v_2,\cdots,v_{n}\}.$$

The vertex set of $G'$ is $V'=V\cup X\cup Y\cup X'\cup Y'$. The subgraph $G'[X]$ (resp. $G'[Y]$) induced by $X$ (resp. $Y$) are $n$ disjoint complete graphs $K_n$, and the subgraph $G'[X']$ (resp. $G'[Y']$) induced by $X'$ (resp. $Y'$) is a complete graph of order $n^3(M+1)$. The subgraph induced by $V$ is clearly the graph $G$. For each $j\in\{1,2,\cdots,n\}$, $v_j$ is connected to both $x_{ij}$ and $y_{ij}$, $1\leq i\leq n$. Moreover, each vertex in $X$ (resp. $Y$) is connected to $n(M+1)$ distinct vertices in $X'$ (resp. $Y'$).

We shall show that $G$ has a dominating set of size at most $k$ if and only if $\kappa^{s}(G',K_{1,M})\leq k$.

First suppose that $G$ has a dominating set $D$ with $|D|\leq k$. We show that there exists a substructure cut $\mathcal{F}$ of $G'$ with $|\mathcal{F}|\leq k$. For any vertex $x\in D$, let $V_x=(N_G(x)\setminus D)\cup\{x\}$. Then $G[V_x]$ is a connected subgraph of $K_{1,M}$ with center vertex $x$. Let $\mathcal{F}=\{G[V_x]|x\in D\}$. Clearly, $V(\mathcal{F})=V$ and $|\mathcal{F}|\leq k$. The removal of $V(\mathcal{F})$ disconnects $G'$ into two components, which implies that $\mathcal{F}$ is a $K_{1,M}$-substructure-cut of size at most $k$. Hence, we have $\kappa^{s}(G',K_{1,M})\leq k$.

Next suppose that $\mathcal{F}=\{F_1,F_2,\cdots,F_k\}$ is a $K_{1,M}$-substructure-cut of $G'$ with $|\mathcal{F}|\leq k$. We show that $G$ has a dominating set of size at most $k$.

Since each element of $\mathcal{F}$ is a connected subgraph of $K_{1,M}$, we focus on the center vertex of $F_i\in\mathcal{F}$ for $1\leq i\leq k$ (since each center vertex of $K_{1,M}$ dominates all its neighbors). Let $S'$ be the set of center vertices of all $F_i\in\mathcal{F}$, $1\leq i\leq k$, and let $S'\cap V=S$. If $S$ is a dominating set of $G$, we are done. Therefore, we assume that $S$ is not a dominating set of $G$. For convenience, we partition $V$ into three disjoint parts $A, B$ and $C$, i.e. $V=A\cup B\cup C$, where $A$ consists of vertices dominated by $S$ in $G$ together with $S$, $B$ consists of vertices that are adjacent to a vertex in $S'\cap (X\cup Y)$, and $C$ is the remaining part.

Thus, two cases arise.

\noindent{\bf Case 1.} $C=\emptyset$. Clearly, $B\neq\emptyset$. Otherwise, $A$ is a dominating set of $G$, a contradiction. So there are vertices of $S'\setminus S$ (lie in $X$ or $Y$) that are adjacent to the vertices of $V(G)$ not dominated by $S$. Observe that the number of vertices of $V(G)$ not dominated by $S$ is not greater than the number of vertices in $S'\setminus S$ since each vertex in $X$ or $Y$ is connected to exactly one vertex in $G$. Consequently, there is a dominating set $S$ of size at most $k$ in $G$.

\noindent{\bf Case 2.} $C\neq\emptyset$. We shall show that $G'-V(\mathcal{F})$ is connected, yielding the desired contradiction.

Since $G'[X']$ is a complete graph of order $n^3(M+1)$, after deleting $V(\mathcal{F})$, i.e. at most $k(M+1)$ (noting that $k(M+1)<n(M+1)$) vertices, the remaining subgraph is still connected.

Note that any vertex $x$ of $X$ not contained in $V(\mathcal{F})$ is connected to exactly $n(M+1)$ neighbors in $X'$. Therefore, any vertex in $X$ (resp. $Y$) of $G'-V(\mathcal{F})$ is connected to vertices in $X'$ (resp. $Y'$). Moreover, any vertex $u\in C$ is connected to $n$ vertices in $X$ (resp. $Y$), at least one of which is not in $V(\mathcal{F})$, so $u$ is connected to vertices of $G'[X']-V(\mathcal{F})$ and $G'[Y']-V(\mathcal{F})$ simultaneously. Similarly, any vertex of $A\cup B$ not in $V(\mathcal{F})$ is connected a vertex in $G'[X']-V(\mathcal{F})$ or $G'[Y']-V(\mathcal{F})$. Thus, $G'-V(\mathcal{F})$ is connected.
%

This completes the proof.
\end{proof}

We state the decision problem of structure connectivity as follows.

\textsc{$H$-Structure Connectivity}

\vskip 0.00 in

\textit{Instance:} Given a nonempty graph $G=(V,E)$, a subgraph $H$ of $G$ and a positive integer $q<|V|$.

\vskip 0.00 in

\textit{Question:} Is $\kappa(G;H)\leq q$?

\begin{corollary}{\bf .}\label{structure}
The \textsc{$H$-Structure Connectivity} is NP-complete when $H=K_{1,3}$.
\end{corollary}
\begin{proof} We adopt all the notations defined in Theorem \ref{substr}. The proof of this corollary differs from that of Theorem \ref{substr} only in the necessity part. If $x\in D$, then $G[V_x]$ may be just a proper subgraph of $K_{1,3}$.
To address this, we augment each $G[V_x]$ by including additional neighbors of $x$ in $X$ or $Y$ into $V_x$ until $G[V_x]$ is isomorphic to $K_{1,3}$. Consequently, we shall obtain $k$ disjoint copies of $K_{1,3}$ as a structure-cut of $G'$.
\end{proof}

\section{$k$-restricted connectivity}

In view of the fact that the NP-completeness of $k$-restricted edge-connectivity, we consider the complexity of restricted connectivity problems in this section. To this end, we present the following two decision problems.

\textsc{$k$-restricted connectivity with three specified vertex sets (RCTSVS)}

\vskip 0.00 in

\textit{Instance:} Given a nonempty graph $G=(V,E)$, a positive integer $k$ and three disjoint vertex subsets $X,Y$ and $Z$ with $\min\{|X|,|Y|,|Z|\}\geq k+1$ and $G[X]$, $G[Y]$ and $G[Z]$ are connected, respectively.

\vskip 0.00 in

\textit{Question:} Is there a set $S$ of vertices ($S\cap (X\cup Y\cup Z)=\emptyset$) such that $S$ separates $X,Y$ and $Z$ (i.e. $X,Y$ and $Z$ lie in different components of $G-S$) and each component of $G-S$ has order at least $k+1$?

\textsc{$R^h$-restricted connectivity with three specified vertex sets}(RhCTSVS).

\vskip 0.00 in

\textit{Instance:} Given a nonempty graph $G=(V,E)$, a positive integer $k$ and three distinct vertex sets $X,Y$ and $Z$ with $\min\{\delta(G[X]),\delta(G[Y]),\delta(G[Z])\}\geq h$.

\vskip 0.00 in

\textit{Question:} Is there a set $S$ of vertices ($S\cap (X\cup Y\cup Z)=\emptyset$) such that $S$ separates $X,Y$ and $Z$ and $\delta(G-S)\geq h$?

To characterize the computational complexity of RCTSVS, we state the following decision problem called \textsc{Non-Monotone 2-3Sat}.

\textsc{Non-Monotone 2-3Sat}

\vskip 0.00 in

\textit{Instance:} Given a set $U$ of variables, a collection $C$ of clauses over $U$ such that each clause $c\in C$ contains two or three variables and for each clause with three variables, it contains at least one negated variable and one un-negated variable.

\vskip 0.00 in

\textit{Question:} Is there a satisfying truth assignment for $C$?

Clearly, \textsc{Non-Monotone 2-3Sat} is a variant of well-known \textsc{Sat}, and its computational complexity is characterized in the following theorem.

\begin{theorem}\cite{Ramarao}{\bf .}\label{2-3SAT}
\textsc{Non-Monotone 2-3Sat} is NP-complete.
\end{theorem}

\begin{theorem}{\bf .}\label{RCTSVS}
RCTSVS is NP-complete for any integer $k\geq1$.
\end{theorem}
\begin{proof} Clearly, the problem is in NP. Given a candidate vertex set $S$, we can verify in polynomial time whether: (1) $G-S$ is disconnected and $X,Y$ and $Z$ lie in distinct components of $G-S$, and (2) each component of $G-S$ has at least $k+1$ vertices. It remains to show that the RCTSVS is NP-hard for any specific integer $k\geq1$. We prove this argument by reducing \textsc{Non-Monotone 2-3Sat} to it.

We first construct a graph $G$ from an instance of \textsc{Non-Monotone 2-3Sat}. Let $U$ be the set of variables and $C$ the collection of clauses. We may assume that $|U|=n$ and there are $m$ clauses of three literals and $p$ clauses of two literals in $C$, that is, $C$ contains $m+p$ clauses in total. For each variable $u_i\in U$, we associate two literal vertices $u_i, \overline{u}_i$ connecting by an edge. For each clause $c_j\in C$, we associate a clause vertex to each literal in it, and label it with that literal, then connect clause vertices each other to form a complete graph of order $|c_j|$. Furthermore, connect each literal vertex to the clause vertex with the same label.

Additionally, create three new vertex sets $X,Y,Z$ of size at least $k+1$, respectively, where $G[X], G[Y]$ and $G[Z]$ are connected, respectively. Connect all literal vertices with labels of un-negated variables to all the vertices in $X$ and all literal vertices with labels of negated variables to all the vertices in $Y$. According to literals in each $c_j$, we consider the following three possibilities:

(1) $c_j$ has two un-negated variables. Connect one of the un-negated variables to all the vertices in $Y$ and the other to all the vertices in $Z$. Clearly, $c_j$ contains one more literal, it must be a negated variable, then connect it to all the vertices in $X$.

(2) $c_j$ has two negated variables. Connect one of the negated variables to all the vertices in $X$ and the other to all the vertices in $Z$. Clearly, $c_j$ contains one more literal, it must be an un-negated variable, then connect it to all the vertices in $Y$.

(3) $c_j$ has exactly one un-negated variable and exactly one negated variable. Connect the negated variable to all the vertices in $X$ and the un-negated to all the vertices in $Y$.

Until now, we have completed the construction of the graph $G$. Fig. \ref{g-NPC-RCTSVS} shows an example of the graph $G$ obtained when $U=\{u_1,u_2,u_3,u_4\}$ and $C=\{\{\overline{u}_1,u_3,\overline{u}_4\},$  $\{u_2,\overline{u}_3,u_4\},\{u_1,\overline{u}_2\}\}$.

\begin{figure}[h]
\centering
\includegraphics[height=70mm]{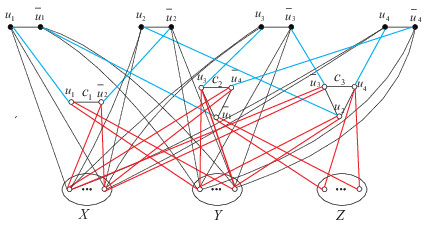}
\caption{An RCTSVS instance constructing from a \textsc{Non-Monotone 2-3Sat} instance in which $U=\{u_1,u_2,u_3,u_4\}$, $C=\{\{u_1,\overline{u}_2\},\{\overline{u}_1,u_3,\overline{u}_4\},\{u_2,\overline{u}_3,u_4\}\}$. Here $s=n+2m+p=9$.}
\label{g-NPC-RCTSVS}
\end{figure}

We claim that if two distinct vertices $u,v\not\in X\cup Y\cup Z$ have the same neighbors in $X$,$Y$ or $Z$, then $u$ and $v$ are not adjacent in $G$. Clearly, this is true if $u$ and $v$ are literal vertices. Similarly, the clause vertices in the same clause are connected to vertices of distinct sets $X$,$Y$ and $Z$ respectively, and the clause vertices in different clauses are non-adjacent. Moreover, two vertices, one literal and one clause, with the same literal label are connected to vertices in distinct independent sets $X$,$Y$ or $Z$, respectively. Thus, the claim holds.

For convenience, let $s=n+2m+p$. In the following, we shall show that there is a satisfying truth assignment for the instance of \textsc{Non-Monotone 2-3Sat} if and only if there is a set $S\subseteq V(G)$ with $|S|=s=n+2m+p$ such that $S$ separates $X,Y$ and $Z$ and each component of $G-S$ is of order at least $k+1$.

{\it Necessity}. Assume that there exists a satisfying truth assignment for the instance of \textsc{Non-Monotone 2-3Sat}. Clearly, for each clause $c_j$, there exists a literal, say $u_i$, with a truth assignment in it. First choose the literal vertices with the same label as $u_i$, and next choose the remaining two or one (noting that the size of $c_j$ is 3 or 2) vertices in $c_j$ whose labels are different from $u_i$. If there exist literal vertices $u_j$ and $\overline{u}_j$ having no truth assignment by the previous operation for some $1\leq j\leq n$, then we add either of $u_j$ and $\overline{u}_j$ but not both to $S$. This generates a set $S$ of vertices of size $s=n+2m+p$. We shall show that the resulting graph $G-S$ is disconnected, each component of $G-S$ has at least $k+1$ vertices, and $X,Y,Z$ lie in different components of $G-S$ respectively.

It is clear that there exist no edges between any two of the sets $X,Y$ and $Z$ in $G$. It can be further deduced that there exists no edges between literal vertices and clause vertices in $G-S$. Therefore, $S$ separates $X,Y$ and $Z$. When a literal vertex or a clause vertex is connected to one vertex in $X$ (or $Y$, $Z$), by the construction of $G$, it is connected to all vertices in $X$, yielding a component of at least $k+1$ vertices containing all vertices of $X$.

{\it Sufficiency}. Suppose that there exists a set $S$ of vertices with $|S|=s=n+2m+p$ such that $S$ separates $X,Y$ and $Z$ and each component of $G-S$ has order at least $k+1$. By the construction of $G$, each literal vertex (resp. clause vertex) is connected to all the vertices in exactly one of $X,Y$ and $Z$.

To separate $X,Y$ and $Z$, it can be deduced from the construction of $G$ that there exists no edges between literal vertices, no edges between literal and clause vertices, and no edges between clause vertices in $G-S$. It follows that for each variable $u_i$, at least one of the literal $u_i$ or $\overline{u}_i$ is contained in $S$. Similarly, at least two vertices of the clause vertices are contained in $S$ for each 3-literal clause, and at least one vertex of the clause vertices is contained in $S$ for each 2-literal clause. By this argument, it follows that $S$ contains at least $s$ vertices. That is, $S$ must in fact contain exactly one of the literal $u_i$ or $\overline{u}_i$, exactly two vertices of the clause vertices for each 3-literal clause, and exactly one vertex of the clause vertices for each 2-literal clause.

If two (or one) clause vertices of a clause $c_j$ are contained in $S$, then the remaining clause vertex, say $u_i$ of $c_j$ is not in $S$. Correspondingly, the corresponding literal vertex with label $u_i$ must be in $S$. Thus, the truth assignment to each literal vertex of $c_j$ not in $S$ is obvious a satisfying truth assignment.

This completes the proof.\end{proof}

\begin{corollary}{\bf .}\label{RhCTSVS}
RhCTSVS is NP-complete for any integer $h\geq1$.
\end{corollary}
\begin{proof} We adopt the notations (including the graph $G$) from Theorem \ref{RCTSVS}. Furthermore, we assume that the minimum degree of each of the subgraphs of $G[X], G[Y]$ and $G[Z]$ is at least $h$. It follows that $|X|\geq h+1, |Y|\geq h+1$ and $|Z|\geq h+1$. Thus, any vertex $u\in V(G)
\setminus (S\cup X\cup Y\cup Z)$ is connected to all the vertices in $X$, $Y$ or $Z$, which implies that $d_{G-S}(u)\geq h$.
Using an argument analogous to that in the proof of Theorem \ref{RCTSVS}, we can deduce that there is a satisfying truth assignment for the instance of \textsc{Non-Monotone 2-3Sat} if and only if there is a set $S\subseteq V(G)$ with $|S|=s=n+2m+p$ such that $S$ separates $X,Y$ and $Z$ and $\delta(G-S)\geq h$.
\end{proof}

To characterize computational complexity of $k$-restricted connectivity of a nonempty graph, we make a step further to state the following two decision problems.

\textsc{$k$-restricted connectivity}($k$-RC)

\vskip 0.00 in

\textit{Instance:} Given a nonempty graph $G=(V,E)$ and a positive integer $k\geq1$.

\vskip 0.00 in

\textit{Question:} Is there a set $S$ of vertices such that $G-S$ is disconnected and each component of $G-S$ has order at least $k+1$?

\textsc{$R^h$-restricted connectivity}(RhC)

\vskip 0.00 in

\textit{Instance:} Given a nonempty graph $G=(V,E)$ and a positive integer $h\geq1$.

\vskip 0.00 in

\textit{Question:} Is there a set $S$ of vertices such that $G-S$ is disconnected and $\delta(G-S)\geq h$?

\begin{theorem}{\bf .}\label{RC}
$k$-RC is NP-complete for any given integer $k\geq1$.
\end{theorem}
\begin{proof} Clearly, the problem is in NP, because we can check in polynomial time whether a set $S$ of vertices such that $G-S$ is disconnected and each component of $G-S$ is of order at least $k+1$. It remains to show that the $k$-RC is NP-hard for any specific integer $k\geq1$. We restrict RCTSVS to $k$-RC by allowing only instances for which $S$ separates three specified distinct vertex sets $X,Y$ and $Z$ with $\min\{|X|,|Y|,|Z|\}\geq k+1$ and $G[X],G[Y],G[Z]$ being connected, respectively, thereby yielding a problem identical to RCTSVS.

%
%

This completes the proof. \end{proof}

\begin{theorem}{\bf .}\label{RhC}
RhC is NP-complete for any given integer $h\geq1$.
\end{theorem}
\begin{proof} Clearly, RhC is in NP, because we can check in polynomial time whether a set $S$ of vertices such that $G-S$ is disconnected and $\delta(G-S)\geq h$.

It remains to show that RhC is NP-hard for any specific integer $h\geq1$. We restrict RhCTSVS to RhC by allowing only instances for which $S$ separates three specified distinct vertex sets $X,Y$ and $Z$ with $\delta(G-S)\geq h$, thereby yielding a problem identical to RhCTSVS.

%
%

This completes the proof. \end{proof}

\section{Conclusions}

In this paper, we give several complexity results concerning variations of graph connectivity, i.e. structure connectivity, substructure connectivity, $k$-restricted connectivity and $R^h$-restricted connectivity. We also note that component connectivity is essentially equivalent to a variant of vertex connectivity (2PCP) studied by Ramarao \cite{Ramarao}, which implies its NP-completeness. In particular, the results show that these network reliability measurements are all NP-complete, which indicates that it is unlikely that there exist polynomially bounded algorithms to solve these problems. Consequently, it is meaningful to study the parameterized complexity, and bounds of these parameters for general graphs, to determine exact values or polynomial-time algorithms for networks with desired properties, and to explore if the problems remain NP-complete on specific graph classes such as interval graphs and chordal graphs.

%

\end{document}